\documentclass{ws-ijmpa}

\def\Journal#1#2#3#4{{#1} {\bf #2}, #3 (#4)}

\def\NIMA{{\em Nucl. Instrum. Methods} A}

\def\PLB{{\em Phys. Lett.}  B}
\def\PRL{\em Phys. Rev. Lett.}
\def\PRD{{\em Phys. Rev.} D}
\def\ZPC{{\em Z. Phys.} C}
\def\EPJC{{\em Eur. Phys. J.} C}

\def\ra{\rightarrow}

\def\be{\begin{equation}}
\def\ee{\end{equation}}
\def\bea{\begin{eqnarray}}
\def\eea{\end{eqnarray}}

\newcommand{\kskl}{K_S^0 K_L^0}
\newcommand{\ks}{K_S^0}

\newcommand{\BR}{{\cal B}}

\newcommand{\psp}{\psi^\prime}
\newcommand{\jpsi}{J/\psi}
\newcommand{\chicJ}{\chi_{cJ}}

\newcommand{\pip}{\pi^+}
\newcommand{\pim}{\pi^-}

\newcommand{\pp}{\pi^+\pi^-}
\newcommand{\kk}{K^+K^-}
\newcommand{\ppb}{p\overline{p}}
\newcommand{\ff}{f_0(980)f_0(980)}
\newcommand{\aab}{\Lambda\overline{\Lambda}}
\newcommand{\ksks}{K^0_S K^0_S}
\newcommand{\kskp}{K^0_S K^+ \pi^- + c.c.}
\newcommand{\kskc}{K^{*+}K^- + c.c.}
\newcommand{\kskn}{K^{*0}\overline {K^0} + c.c.}

\newcommand{\jpsito}{J/\psi \rightarrow }

\newcommand{\pspto}{\psp \rightarrow }
\newcommand{\chicJto}{\chi_{cJ} \rightarrow }
\newcommand{\chiczto}{\chi_{c0} \rightarrow }
\newcommand{\chicoto}{\chi_{c1} \rightarrow }
\newcommand{\chictto}{\chi_{c2} \rightarrow }
\newcommand{\bfg}{\begin{figure}}
\newcommand{\efg}{\end{figure}}
\newcommand{\bitm}{\begin{itemize}}
\newcommand{\eitm}{\end{itemize}}
\newcommand{\bnum}{\begin{enumerate}}
\newcommand{\enum}{\end{enumerate}}
\newcommand{\btbl}{\begin{table}}
\newcommand{\etbl}{\end{table}}
\newcommand{\btbu}{\begin{tabular}}
\newcommand{\etbu}{\end{tabular}}

\begin{document}

\markboth{C.~Z.~Yuan} {Hadronic Decays of Charmonia from BESII}

%
\catchline{}{}{}{}{}
%

\title{Hadronic Decays of Charmonia from BESII}

\author{\footnotesize Chang-Zheng~Yuan (For the BES Collaboration)}

\address{Institute of High Energy Physics, Chinese Academy of Sciences,\\
Beijing 100039, China}

\maketitle

\pub{Received (Day Month Year)}{Revised (Day Month Year)}

\begin{abstract}

Recent results on charmonia decays at BESII/BEPC are reported,
including the observation of $\pspto \kskl$, $\pspto Vector\, \,
Tensor$ and $\pspto Vector\, \, Pseudoscalar$ for the measurement
of the relative phase between the strong and electromagnetic
decays of $\psp$ and a test of the pQCD ``12\% rule'' between
$\psp$ and $\jpsi$ decays; the test of the color-octet mechanism
via $\chicJto \ppb$ and $\chicJto \aab$; the first observation of
$\chiczto \ff$; and a study of the $\psp$ and $\chicJ$ decays with
$\ksks$ in the final states. \keywords{Charmonium; hadronic
decays; pQCD.}
\end{abstract}

\section{BES experiment and the data samples}

The data samples used for the analyses are taken with the Beijing
Spectrometer (BESII) detector~\cite{bes,bes2} at the Beijing
Electron-Positron Collider (BEPC) storage ring at a center-of-mass
energies corresponding to $M_{\psp}$ and $M_{\jpsi}$. The data
samples contain $(14 \pm 0.6)\times 10^6$ $\psp$ events and $(57.7
\pm 2.7)\times 10^6$ $\jpsi$ events, as determined from inclusive
hadronic decays.

\section{Observation of $\pspto \kskl$}

It has been determined that for many two-body exclusive $\jpsi$
decays~\cite{suzuki,jphase,wymphase} the relative phases between
the three-gluon and the one-photon annihilation amplitudes are
near $90^\circ$.  For $\psp$ decays, the available information
about the phase is much more limited because there are fewer
experimental measurements. The analysis of $\pspto Vector \,\,
Pseudoscalar$ (VP) decays shows that the phase could be the same
as observed in $\jpsi$ decays~\cite{wymphase}, but it could not
rule out the possibility that the phase is near $180^\circ$ as
suggested in Ref.~\cite{suzuki} due to the big uncertainties in
the experimental data. A measurement of the relative phase in
$\pspto Pseudoscalar\,\, Pseudoscalar$ (PP) is suggested in
Ref.~\cite{phase_pp} by searching for $\pspto \kskl$.

\begin{figure}[htbp]
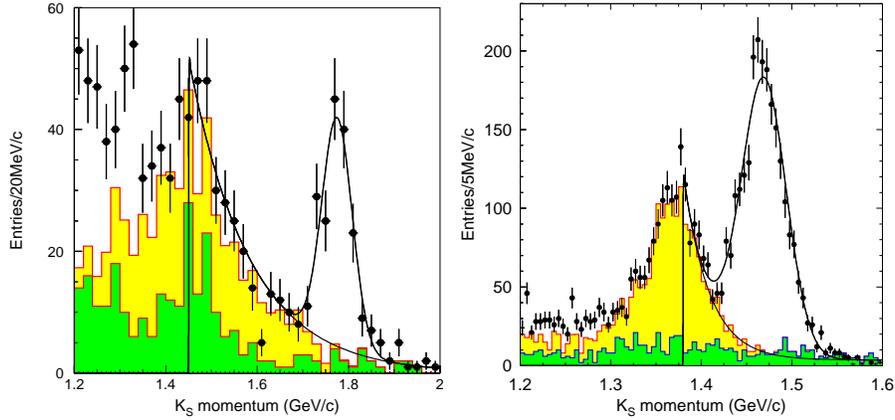

\centerline{\hbox{
\psfig{file=pksdtfit_prl_new.epsi,height=5.5cm}} \hbox{
\psfig{file=pks_fit_j_prd.epsi,height=5.5cm}}} \caption{The $K_S$
momentum distribution for data at $\psp$ (left) and $\jpsi$
(right). The dots with error bars are data and the curves are the
best fit of the data. The dark shaded histogram is from $K_S$ mass
side band events, and the light shaded histogram is from the Monte
Carlo simulated backgrounds. } \label{kskl}
\end{figure}

BESII searches for $\pspto \kskl$ by reconstructing the monochroic
$\ks$ in the 14~M $\psp$ data sample~\cite{bes2kskl}. The signal,
as shown in Fig.~\ref{kskl}, is very significant (about
13$\sigma$), and the branching fraction is measured to be \(
\BR(\pspto \kskl) = (5.24\pm 0.47 \pm 0.48)\times 10^{-5}\). This
branching fraction, together with branching fractions of $\pspto
\pp$ and $\pspto \kk$, are used to extract the relative phase
between the three-gluon and the one-photon annihilation amplitudes
of the $\psp$ decays to pseudoscalar meson pairs. It is found that
a relative phase of $(-82\pm 29)^{\circ}$ or $(+121\pm
27)^{\circ}$ can explain the experimental results~\cite{phase_pp}.

A similar analysis of the $\jpsi$ data sample yields an improved
measurement of the $\jpsito \kskl$ (see Fig.~\ref{kskl}) branching
fraction~\cite{bes2ksklj}: \( \BR(\jpsito \kskl) = (1.82\pm 0.04
\pm 0.13)\times 10^{-4}\), which is more than $4\sigma$ larger
than the world average~\cite{pdg}. Comparing with the
corresponding branching fraction for $\pspto \kskl$ , one gets
      \( Q_h = \frac{\BR(\pspto \kskl)}{\BR(\jpsito \kskl)}
                   = (28.8\pm 3.7)\% \).
This result indicates that $\psp$ decays is enhanced by more than
4$\sigma$ relative to the ``12\% rule'' expected from perturbative
QCD (pQCD)~\cite{beswangwf}, while for almost all other channels
where the deviations from the ``12\% rule'' are observed, $\psp$
decays are suppressed.

The violation of the ``12\% rule'' in $\kskl$ mode is explained in
Ref.~\cite{wmykskl} in the $S$- and $D$-wave mixing model of the
$\psp$ state. In this scenario, the $\psi(3770)$, also an $S$- and
$D$-wave mixed charmonium state will have a decay branching
fraction to $\kskl$ between $(0.12\pm 0.07)\times 10^{-5}$ and
$(3.8\pm 1.1)\times 10^{-5}$. This need to be tested with the
large $\psi(3770)$ data samples at CLEOc and BESIII.

\section{Observation of $\pspto Vector \,\, Tensor$}

Four $Vector \,\, Tensor$ (VT) decay channels $\pspto \omega
f_{2}(1270) \rightarrow \pi^+\pi^-\pi^+\pi^-\pi^0$, $\rho
a_2(1320) \rightarrow \pi^+\pi^-\pi^+\pi^-\pi^0$,
$K^*(892)^0\overline{K}^*_2(1430)^0+c.c. \ra \pi^+\pi^-K^+K^-$ and
$\phi f_2^{\prime}(1525) \ra K^+K^-K^+K^-$ are investigated to
test the pQCD ``$12\%$ rule''~\cite{beswangwf}. Previous BESI
results~\cite{BESVT,beswangwf} on these channels reveal that these
VT decay modes are suppressed compared to the pQCD prediction.
However, the measurements, using about $4\times 10^6$ $\psp$
events, determined only upper limits or branching fractions with
large errors. These analyses are updated with $14\times 10^6$
$\psp$ events, and signals of all these four channels are
observed~\cite{bes2vt}.  The statistical significance for all four
channels are larger than $3 \sigma$; those for $\omega f_2(1270)$
and $K^*(892)^0\overline{K}^*(1430)^0+c.c.$ are larger than $5
\sigma$. Table~\ref{BESII_VT} summarizes the results of the four
branching fraction measurements, as well as the corresponding
branching fractions of $J/\psi$ decays, and the ratios of the
$\psp$ to $J/\psi$ branching fractions. All four VT decay modes
are suppressed by a factor of 3 to 5 compared with the pQCD
expectation.

\begin{table}
\tbl{\label{BESII_VT}
  Branching fractions measured for
  $\pspto Vector\,\, Tensor$. Results for corresponding
  $J/\psi$ branching fractions are also given as well as the
  ratio $Q_X=\frac{\BR(\pspto X)}{\BR(\jpsito X)}$. }
{\begin{tabular}{@{}ccccc@{}} \toprule
 X &$N^{obs}$ &  $B(\psp\ra
X)(\times
10^{-4})$ & $B(J/\psi\ra X)(\times 10^{-3})$ & $Q_X(\%)$ \\
\hline $\omega f_{2}$  & $62\pm12$  & $2.05\pm0.41\pm0.38$
 & $4.3\pm0.6$ & $4.8\pm 1.5$ \\
$\rho a_2$              & $112\pm31$
& $2.55\pm0.73\pm0.47$ & $10.9\pm2.2$ & $2.3\pm1.1$ \\
$K^{*0}\overline{K}^{*0}_2+c.c.$   & $93\pm16$
& $1.86\pm0.32\pm0.43$ & $6.7\pm2.6$  &  $2.8\pm1.3$ \\
$\phi f_2^{\prime}$     & $19.7\pm5.6$  & $0.44\pm0.12\pm0.11$
 & $1.23\pm 0.21$ & $3.6\pm1.5$ \\\botrule
\end{tabular}}
\end{table}

\section{Observation of $\pspto Vector \,\, Pseudoscalar$ (preliminary)}

This mode is also investigated to test the pQCD ``$12\%$
rule''~\cite{beswangwf} and to study the relative phase between
strong and electromagnetic decays of the
charmonium~\cite{wymphase}. Here $\kskc$ and $\kskn$ are studied
with $\kskp$ final state.

In the event, $\ks$ is identified through the decay $\ks\ra
\pip\pim$. Any two oppositely charged tracks are assumed to be
pions and their intersection is found with the code used in
Ref.~\cite{bes2kskl}, a $\ks$ candidate should have $\pip\pim$
invariant mass agree with $\ks$ nominal mass and the decay length
in the transverse plane ($L_{xy}$) should be large. The four
charged tracks are kinematically fitted to further improve the
resolution and remove more background.

\begin{figure}[htb]
\centerline{\hbox{ \psfig{file=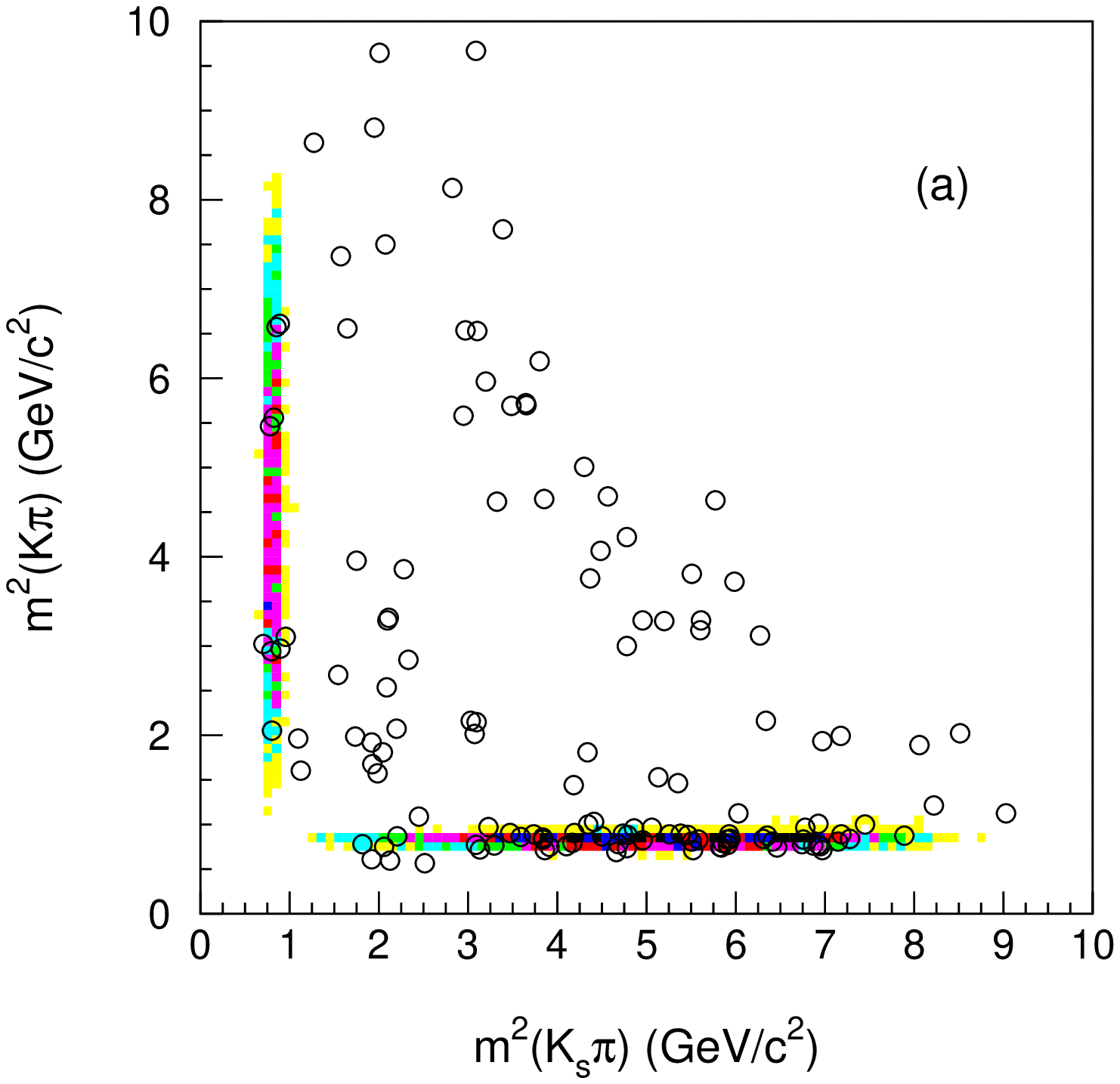,height=6.5cm}} \hbox{
\psfig{file=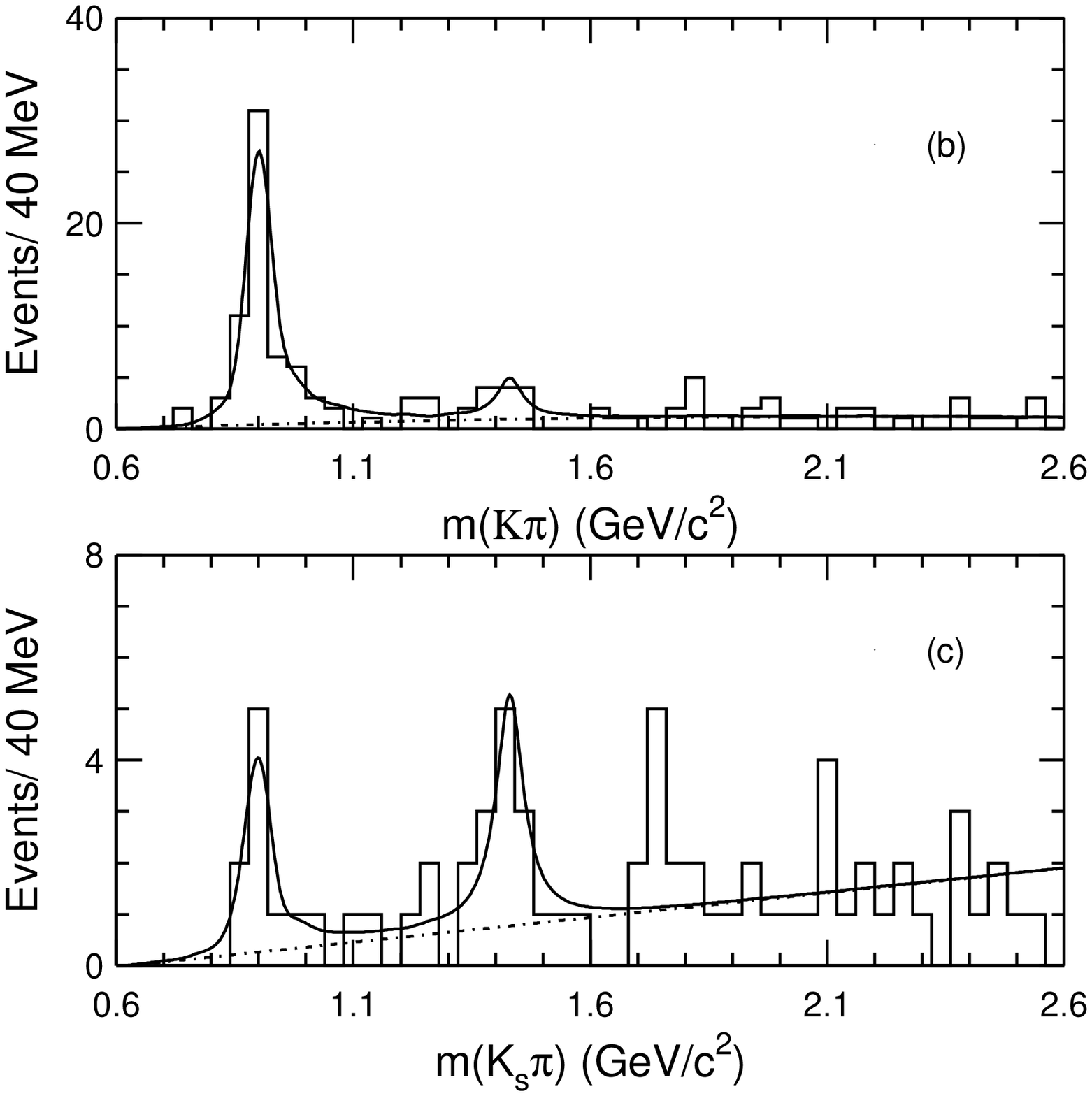,height=6.5cm}}}
\caption{\label{k3pi_Dalitz} Dalitz plot (a) and invariant mass of
$K^{\pm}\pi^{\mp}$ (b) and of $\ks\pi^{\pm}$ (c) for $K_s^0
K^\pm\pi^\mp$ candidate events after cuts described in the text.
The blank dot stands for real data, and the shadow for Monte Carlo
simulation of $\pspto K^*\bar{K}+c.c.$ (the horizontal cluster for
$\kskn$ and the vertical cluster for $\kskc$). The curve shows the
best fit described in the text. }
\end{figure}

After above selection, the Dalitz plot of the candidate events is
shown in Fig.~\ref{k3pi_Dalitz}(a). Monte Carlo simulation of
$\pspto K^{*}\bar{K}+c.c.$ indicates that $\kskn$ appears as a
horizontal band and $\kskc$ as a vertical band.
Figure~\ref{k3pi_Dalitz}(b) shows the invariant mass of
$K^{\pm}\pi^{\mp}$ with an additional cut
$m_{\ks\pi^{\pm}}>1.0$~GeV to remove the influence from $\kskc$,
and Figure~\ref{k3pi_Dalitz}(c) shows the invariant mass of
$\ks\pi^{\pm}$ for $\ks K^{\pm}\pi^{\mp}$ candidate events with an
additional cut $m_{K^{\pm}\pi^{\mp}}>1.0$~GeV to remove the
influence from $\kskn$. Clear $\kskc$ and $\kskn$ signals are
observed in the plots.

The fit of the invariant mass spectra for $K^{\pm}\pi^{\mp}$ and
$\ks\pi^{\mp}$, as shown in Fig.~\ref{k3pi_Dalitz} (b) and (c),
yield $65.6\pm9.0$ $\kskn$ and $9.6\pm4.2$ $\kskc$ events. Their
detection efficiencies are $(9.68\pm0.07)\%$ and $(7.25\pm0.07)\%$
from Monte Carlo simulation, and their statistical significance
are $11\sigma$ and $3.5\sigma$, respectively. The branching
fractions are listed in Table~\ref{BESII_VP}, compared with the
corresponding branching fractions of $\jpsi$ decays, $\psp$ decays
are suppressed relative to the 12\% rule expectation.

\begin{table}
\tbl{ \label{BESII_VP} Branching fractions measured for $\pspto
K^{*}\bar{K}+c.c.$ (preliminary). Results for corresponding
$J/\psi$ branching fractions by PDG are also given as well as the
ratios $Q_h=\frac{B(\psp)}{B(J/\psi)}$. }
{\begin{tabular}{@{}cccc@{}} \toprule Channels & $\BR(\psp)$
($\times 10^{-5}$) & $\BR(J/\psi)$ ($\times 10^{-4}$)& $Q_h$ (\%)
\\\colrule
$\kskc$  & $2.9\pm1.4\pm0.4$ & $50\pm4$ & $0.58\pm0.29$\\
$\kskn$  & $15.0\pm2.1\pm1.7$ & $42\pm4$ & $3.6\pm0.7$
\\\botrule
\end{tabular}}
\end{table}

The ratio $\frac{\BR(\kskn)}{\BR(\kskc)} =5.1\pm2.5$ shows a large
isospin violation between the charged mode and neutral mode of
$\pspto K^{*}\bar{K}+c.c.$ decays, this is understandable since
the electromagnetic decay amplitudes are different in these two
channels.

Contributions from the continuum $e^+ e^- \rightarrow
\gamma^*\rightarrow$ hadrons \cite{WYMZ,wymphase} are estimated
using a data sample of $(6.42\pm0.24)$~pb$^{-1}$ taken at $\sqrt
s=3.65$~GeV, about one-third of the integrated luminosity at the
$\psp$. $2.5\pm1.9$ events are observed in $K^*(892)^0
\bar{K}^0+c.c.$, while no events in charged channel. Since the
signals are not significant due to the limited statistics, they
are not considered in the branching ratio determination above.
However, if one assumes the continuum amplitude is indeed at the
measured level, the branching fractions of $\pspto \kskc$ and
$\kskn$ can be recalculated by the model proposed in
Ref.~\cite{wymphase}, where the contributions of the continuum and
the interference are taken into consideration. The branching
fractions are changed to: $\BR(\pspto\kskc)=3.1\times 10^{-4}$ and
$\BR(\pspto\kskn)=12.7\times 10^{-4}$, where the uncertainties due
to the model are not included.

 In conclusion, we present the branching fractions  for
$\pspto \kskn$ and $\kskc$ for the first time, they are suppressed
with respect to the pQCD expectation, and a large isospin
violation in the charged and neutral mode is observed. This may
shed light on the understanding the $\psp$ decay dynamics.

\section{Test of COM in P-wave charmonium Baryonic decays}

Hadronic decay rates of P-wave quarkonium states provide good
tests of QCD. The decays $\chicJto \ppb$ have been calculated
using different models~\cite{besi5}, and recently, the decay
branching fractions of $\chicJto$ baryon and anti-baryon pairs
were calculated including the contribution of the color-octet fock
states~\cite{wong}.  Using the $\chicJto \ppb$ branching fractions
as input to determine the matrix element, the partial widths of
$\chicJto \aab$ are predicted to be about half of those of
$\chicJto \ppb$, for $J=1$ and $2$.  As shown in Table~\ref{br},
the measurements of $\chicJto \aab$~\cite{aa} together with the
branching fractions of $\chicJto \ppb$~\cite{bes2ppb} from the
same data sample, indicate that $\chicJto \aab$ is enhanced
relative to $\chicJto \ppb$, as compared with the color-octet
mechanism (COM) calculation~\cite{wong}.

\begin{table}[htbp]
\tbl{Branching fractions of $\chicJto \aab$ and $\chicJto \ppb$,
and \( R_{\cal B} = \BR(\chicJto \aab)/\BR(\chicJto \ppb).\)}
{\begin{tabular}{@{}cccc@{}} \toprule

$\BR(\chicJto \aab)$ ($10^{-5}$) &$47^{+13}_{-12}\pm10$
                                &$26^{+10}_{-9}\pm 6$
                                &$33^{+15}_{-13}\pm 7$  \\
$\BR(\chicJto \ppb)$ ($10^{-5}$)&$27.1^{+4.3}_{-3.9}\pm4.7$
                                &$5.7^{+1.7}_{-1.5}\pm0.9$
                                &$6.5^{+2.4}_{-2.1}\pm1.0$  \\
$R_{\cal B}$&$1.73\pm0.63$&$4.6\pm2.3$&$5.1\pm3.1$\\
\botrule
\end{tabular}}
\label{br}
\end{table}

\section{Evidence for $\chiczto \ff$}

After thirty years of controversy, the nature of the $f_0(980)$ is
still not settled~\cite{REVIEW,pdg}, and more experimental results
are needed to clarify it. Here we report on the analysis of
$\pi^+\pi^-\pi^+\pi^-$ final states from $\chi_{c0}$ decays using
the $\psp$ data sample. Evidence for $f_0(980)f_0(980)$ production
from $\chi_{c0}$ decays is obtained for the first
time~\cite{besf0f0}.

\begin{figure}[htb]
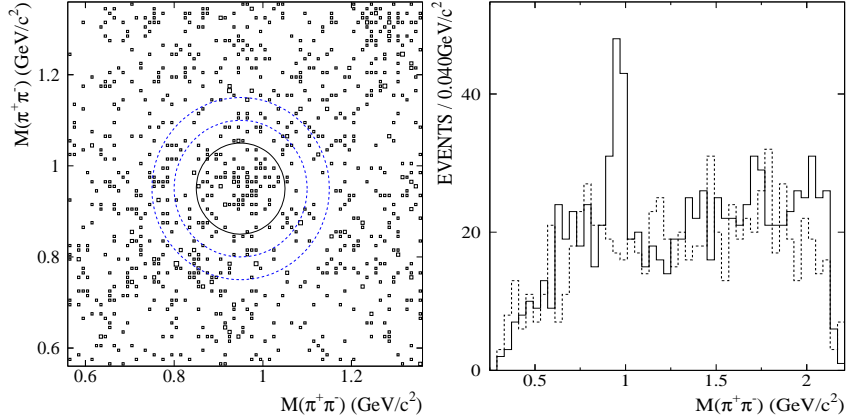

\centerline{\hbox{\psfig{file=scaa.epsi,height=5.5cm,width=5.5cm}}
\hbox{\psfig{file=f0-n.epsi,height=5.5cm,width=5.5cm}}}
\caption{Left: scatter plot of $\pi^+ \pi^-$ versus $\pi^+ \pi^-$
invariant mass in the $f_0(980)$ region for $\chi_{c0}$ candidate
events, the circles show the definition of signal and background
regions. Right: $\pi^+\pi^-$ mass distribution recoiling against
the $f_0(980)~(0.88$~GeV $ < m_{\pi^+\pi^-} < 1.04 $~GeV) for
events in the $\chi_{c0}$ mass region (two entries per event),
where the dashed line histogram indicates a rough estimation of
background determined from sidebands.} \label{scaa}
\end{figure}

The left plot of Figure~\ref{scaa} shows scatter plot of
$\pi^+\pi^-$ versus $\pi^+\pi^-$ invariant mass for the events in
$\chi_{c0}$ mass region (from 3.30 to 3.48~GeV), and the
definition of the signal and background control regions. The
signal region is shown in the figure as a circle centered at
$(0.960, 0.960)$~GeV and with a radius of 80~MeV, and the
background is estimated from the events between two circles with
radii of 120~MeV and 160~MeV. There are 65 and 51 events in the
signal and background regions, respectively. So the number of
$f_0(980)f_0(980)$ events is estimated to be 65 - 51/1.75 =
$35.9\pm 9.0$, where 1.75 is the normalization factor -- the ratio
of the area of background region to that of the signal region. We
obtain the signal significance of the $f_0(980)f_0(980)$ of
4.6$\sigma$ using the method described in Ref.~\cite{Narsky}.
After requiring that the mass of one of the $\pi^+\pi^-$ pairs
lies between 0.88 and 1.04~GeV, the mass distribution of the other
$\pi^+\pi^-$ pair is shown in the right plot of Fig.~\ref{scaa}
(two entries per event); there is a strong $f_0(980)$ signal, and
its line shape is similar to other experiments~\cite{pdg}.

The resulting branching ratio is
$${\cal B}(\psp\to\gamma\chi_{c0}\to \gamma f_0(980)f_0(980)\to
\gamma\pi^+\pi^-\pi^+\pi^-) = (6.5\pm 1.6\pm 1.3)\times 10^{-5},$$
and using the PDG~\cite{pdg} average value and error for ${\cal
B}(\psp\to\gamma\chi_{c0})$, we obtain
$${\cal B}(\chi_{c0}\to f_0(980)f_0(980)\to\pi^+\pi^-\pi^+\pi^-) = (7.6\pm
1.9~(\mbox{stat})\pm 1.6~(\mbox{syst}))\times 10^{-4}.$$ This may
help in understanding the nature of $f_0(980)$.

\section{Search for $\psp$ and $\jpsito \ksks$}

The CP violating processes $\jpsito \ksks$ and $\pspto \ksks$ are
searched for using the $\jpsi$ and $\psp$ samples~\cite{bes2ksks}.
One candidate in each case is observed, in agreement with the
expected background level. The upper limits on the branching
ratios are determined to be \( \BR(\jpsito \ksks) <1.0\times
10^{-6} \) and \( \BR(\pspto \ksks) <4.6\times 10^{-6} \) at the
95\% C. L. The former is much more stringent than the previous
Mark-III measurement~\cite{mk3ksks}, and the latter is the first
search for this channel in $\psp$ decays. The current bounds on
the production rates are still far beyond the sensitivity needed
for testing the EPR paradox~\cite{EPR}, and even farther for CP
violation~\cite{roo}.

\section{$\psp$ and $\chicJ$ decays with $\ksks$ in the final states (preliminary)}

Since the observed total branching fractions of $\psp$ and
$\chicJ$~($J=0,1,2$) decays into light hadrons are still small
(about $1.5\sim 11\%$\cite{pdg}). The final states with two
$K^0_S$ are searched for using the $\psp$ data sample. The
analyses including $\pspto \pip \pim \ksks$, $\chicJto \ks\ks$,
$\pip \pim \ksks$ and $K^+K^- \ksks$, except for $\chicJto \ks\ks$
which has been measured by BES experiment~\cite{bes1ksks}, other
channels were not observed before.

\begin{figure}[htb]
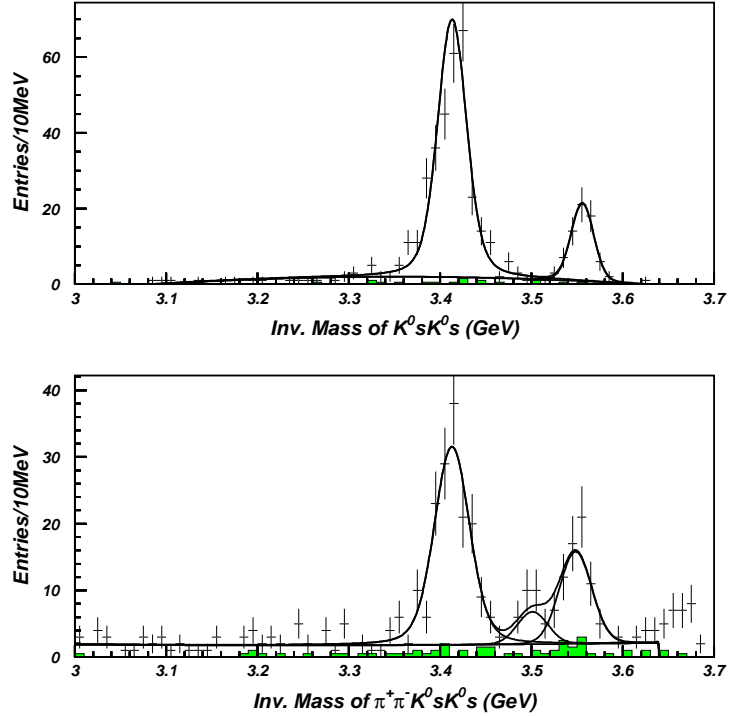

\centerline{\psfig{figure=r2ks.epsi,width=9.5cm}}\bigskip
\centerline{\psfig{figure=r2pi2ks.epsi,width=9.5cm}}
\caption{Invariant mass of $K^0_SK^0_S$ in $\psp\rightarrow \gamma
K^0_SK^0_S$ (top) and invariant mass of $\pi^+\pi^-K^0_SK^0_S$ in
$\psp\rightarrow \gamma\pi^+\pi^-K^0_SK^0_S$ (bottom). The error
bars are data and the shaded histograms are background estimated
from the $K^0_S$ mass sidebands. The solid lines are the best fit
to the data.} \label{wangzhe}
\end{figure}

Figure~\ref{wangzhe} shows the invariant mass of $\ks\ks$ in
$\gamma \ks \ks$ channel, and that of $\pip \pim \ksks$ in $\gamma
\pip \pim \ksks$ channel. Clear $\chicJ$ states can be seen. Final
results of branching fractions are summarized in
Table~\ref{tab:fitting}. The branching fractions of
$\chi_{cJ}$~(J=0,1,2) decays to $\pi^+\pi^-K^0_SK^0_S$ and
$K^+K^-K^0_SK^0_S$, as well $\psp$ decay to $\pi^+\pi^-K^0_SK^0_S$
are measured for the first time. The branching fractions of
$\chi_{c0}$ and $\chi_{c2}$ decays to $K^0_SK^0_S$ are measured
with improved precision.

\begin{table}
\tbl{Summary of the branching ratio results (preliminary). The
first and second errors for branching fractions are statistical
and systematic respectively.} {\begin{tabular}{@{}ccc@{}} \toprule
                        Channel & $n^{obs}$ & Branching Ratio \\\colrule
$\chiczto \ks\ks$&{$302\pm18$}&{$(3.37\pm0.20\pm0.44)\times10^{-3}$}\\
$\chicoto \ks\ks$&{0}         &{$<2.55\times10^{-5}$ (CL=90\%)}\\
$\chictto
\ks\ks$&{$64.8\pm8.0$}&{$(8.9\pm1.1\pm1.2)\times10^{-4}$}\\\colrule
$\chiczto \pi^+\pi^-\ks\ks$&{$157\pm12$}&{$(6.6\pm0.5\pm1.1)\times10^{-3}$}\\
$\chicoto \pi^+\pi^-\ks\ks$&{$20.4\pm6.9$}&{$(8.0\pm3.1\pm1.3)\times10^{-4}$}\\
$\chictto
\pi^+\pi^-\ks\ks$&{$62.2\pm9.2$}&{$(3.24\pm0.57\pm0.55)\times10^{-3}$}\\\colrule
$\chiczto K^+K^-\ks\ks$&{$19.2\pm3.7$}&{$(1.87\pm0.36\pm0.34)\times10^{-3}$}\\
$\chicoto K^+K^-\ks\ks$&{$3.9\pm2.4$}&{$(3.1\pm1.9\pm0.6)\times10^{-4}$}\\
$\chictto
K^+K^-\ks\ks$&{$3.0\pm2.1$}&{$(3.3\pm2.3\pm0.6)\times10^{-4}$}\\\colrule
$\pspto \pi^+\pi^-\ks\ks$&{$83.2\pm4.4$}&{$(2.20\pm0.12\pm0.33)\times10^{-4}$}\\
\botrule
\end{tabular}}
\label{tab:fitting}
\end{table}

\section*{Acknowledgments}

Thanks my colleagues of BES collaboration who did the good works
which are reported here, and thanks the organizers for the kind
invitation and the successful organization of the workshop.


\begin{thebibliography}{0}

\bibitem{bes} J.~Z.~Bai. {\em et al.} (BES Collab.), \Journal\NIMA{344}{319}{1994}.

\bibitem{bes2} J.~Z.~Bai. {\em et al.} (BES Collab.),\Journal\NIMA{458}{627}{2001}.

\bibitem{suzuki} M.~Suzuki, \Journal\PRD{63}{054021}{2001};
                 J.~L.~Rosner, {\em ibid.} {\bf 60}, 074029 (1999).

\bibitem{jphase} J.~Jousset {\em et al.},
\Journal\PRD{41},{1389}{1990};
 D.~Coffman {\em et al.}, {\em ibid.} {\bf 38}, {2695} {(1988)};
 M.~Suzuki, {\em ibid.} {\bf 60}, {051501}, {(1999)};
 L.~K\"{o}pke and N.~Wermes,
                 {\em Phys. Rep.} {\bf 174}, {67} {(1989)};
 R.~Baldini {\em et al.}, \Journal\PLB{444}{111}{1998}.

\bibitem{wymphase} P.~Wang, C.~Z.~Yuan and X.~H.~Mo,
                  \Journal\PRD{69}{057502}{2004}.

\bibitem{phase_pp} C.~Z.~Yuan, P.~Wang and X.~H.~Mo,
                   \Journal\PLB{567}{73}{2003}.

\bibitem{bes2kskl} J.~Z.~Bai. {\em et al.} (BES Collab.),
              \Journal\PRL{92}{052001}{2004}.

\bibitem{bes2ksklj} J.~Z.~Bai. {\em et al.} (BES Collab.),
              \Journal\PRD{69}{012003}{2004}.

\bibitem{pdg} S.~Eidelman {\em et al.} (Particle Data Group),
              \Journal\PLB{592}{1}{2004}.

\bibitem{beswangwf} J.~Z.~Bai. {\em et al.} (BES Collab.),
             \Journal\PRD{67}{052002}{2003}.

\bibitem{wmykskl}P.~Wang, X.~H.~Mo and C.~Z.~Yuan, hep-ph/0402227.

\bibitem{BESVT} J. Z. Bai {\em et al.} (BES Collab.),
      \Journal\PRL{81}{5080}{1998}.

\bibitem{bes2vt} J. Z. Bai {\em et al.} (BES Collab.), \Journal\PRD{69}{072001}{2004}.

\bibitem{WYMZ} P. Wang, C. Z. Yuan, X. H. Mo and D. H. Zhang,
               \Journal\PLB{593}{89}{2004}.

\bibitem{besi5} M.~Anselmino, R.~Cancelliere, and F.~Murgia,
                 \Journal\PRD{46}{5049}{1992};
                 M.~Anselmino, F.~Caruso, and S.~Forte,
                 {\em ibid.} {\bf 44}, 1438 (1991);
               M.~Anselmino and F.~Murgia, \Journal\ZPC{58}{429}{1993}.

\bibitem{wong} S.~M.~H.~Wong, \Journal\EPJC{14}{643}{2000}.

\bibitem{aa} J.~Z.~Bai. {\em et al.} (BES Collab.),
              \Journal\PRD{67}{112001}{2003}.

\bibitem{bes2ppb} J.~Z.~Bai. {\em et al.} (BES Collab.),
              \Journal\PRD{69}{092001}{2004}.

\bibitem{REVIEW} For general reviews, see S. Godfrey and J. Napolitano,
{\em Rev. Mod. Phys.} {\bf 71}, 1411 (1999); C. Amsler and N.
T\"{o}rnqvist, {\em Phys. Rept.} {\bf 389}, 61 (2004).

\bibitem{besf0f0} M.~Ablikim {\em et al.} (BES Collab.),
              hep-ex/0406079.

\bibitem{Narsky} S.I. Bityukov, JHEP 0209:060, 2002,
   \Journal\NIMA{502}{795}{2003}.

\bibitem{bes2ksks} J.~Z.~Bai. {\em et al.} (BES Collab.),
              \Journal\PLB{589}{7}{2004}.

\bibitem{mk3ksks} R.~M.~Baltrusaitis {\em et al.},
                \Journal\PRD{32}{566}{1985}.

\bibitem{EPR} A.~Einstein, B.~Podolsky and N.~Rosen,
               {\em Phys. Rev.} {\bf 47}, 777 (1935).

\bibitem{roo} M.~Roos, ``Test of Einstein Locality'',
              HU-TFT-80-5 (revised), Nov. 1980.

\bibitem{bes1ksks} J.~Z.~Bai. {\em et al.} (BES Collab.),
              \Journal\PRD{60}{072001}{1999}.

\end{thebibliography}
\end{document}